\documentclass[twoside,twocolumn,english]{revtex4}
\usepackage[T1]{fontenc}
\usepackage[latin1]{inputenc}
\usepackage{amssymb}

\makeatletter

\usepackage{babel}
\makeatother
\begin{document}

\title{Obtention of a Gravitational Force from a Relativistic Solution to
Binet's Equation: The Schwarzschild's Case}

\author{Terenzio Soldovieri%
\footnote{E-mail: tsoldovi@luz.ve%
} \ \& \'{A}ngel G. Mu\~{n}oz S.%
\footnote{E-mail: agmunoz@luz.ve%
}}

\address{Grupo de Investigaciones de Física Teórica (G.I.F.T.) Depto de F\'{\i}sica.
Facultad de Ciencias. \\
 La Universidad del Zulia (LUZ). Maracaibo 4001 -Venezuela.}

\begin{abstract}
Making use of the classical Binet\'{}s equation a general procedure
to obtain the gravitational force corresponding to an arbitrary 4-dimensional
spacetime is presented. This method provides, for general relativistic
scenarios, classics expressions that may help to visualize certain
effects that Newton\'{}s theory can not explain. In particular, the
force produced by a gravitational field which source is spherically
symmetrical (Schwarzschild's spacetime) is obtained. Such expression
uses a redefinition of the classical reduced mass, in the limit case
it can be reduced to Newton's Universal Law of Gravitation and it
produces \textit{two} different orbital velocities for test particles
that asimptotically coincide with the Newtonian one. 



Se presenta, haciendo uso de la ecuaci\'{o}n de Binet cl\'{a}sica,
un procedimiento general para obtener la expresi\'{o}n de la fuerza
gravitacional correspondiente a un espacio-tiempo tetradimensional
arbitrario. Este m\'{e}todo provee expresiones cl\'{a}sicas para
escenarios relativistas, lo que podr\'{\i}a ayudar a visualizar efectos
que no pueden ser explicados por la teor\'{\i}a newtoniana. En particular,
se obtiene la fuerza producida por un campo gravitacional cuya fuente
es esf\'{e}ricamente sim\'{e}trica (espaciotiempo de Schwarzschild).
Tal expresi\'{o}n emplea una redefinici\'{o}n de la \textit{masa
reducida} cl\'{a}sica, en el caso l\'{\i}mite se reduce a la Ley
de Gravitaci\'{o}n Universal de Newton y es similar a otras expresiones
obtenidas en a\~{n}os recientes. 


\textbf{Palabras Clave:} Ley de Gravitaci\'{o}n Universal, corrimiento
del perihelio, potencial de Schwarzschild, masa reducida. 
\end{abstract}
\maketitle

\section{A Historical Introduction}

The obtention of an expression for the gravitational force was first
published in 1687 in the well known book Philosophiae Naturalis Principia
Mathematica \cite{newton}, in which Newton, using his three laws,
deduces \cite{newton2} an expression according to which the intensity
of gravitational attraction is proportional to the product of the
interacting masses and decreases to the square of the distance between
them. In vectorial notation this expression can be written as: \begin{equation}
\overrightarrow{F}=-G\frac{Mm}{r^{2}}\hat{u}_{r}\label{Fu}\end{equation}
 where $G$ is the constant of universal gravitation, $M$ and $m$
the source and the test particle masses respectively, $r$ the distance
separating the bodies and $\hat{u}_{r}$ a versor in radial direction.
As everyone knows, with this set of laws Newton satisfactorily explained
the movement of planets and the other celestial bodies, giving, in
this way, the foundations of Modern Astronomy. 

Even though Newton's gravitation theory gave a good explanation of
orbital phenomena happening in the heavens, with time, observations
demonstrated that certain discrepancies that could not be adequately
explained did exist. After Newton, Laplace and Poisson rewrote the
gravitational law, giving it a both mathematical and physical more
elegant formalism, but which still did not solve the aforesaid discrepancies.
In fact, in 1895 Simon Newcomb, after numerous attempts to find a
solution to the problem, suggested that perhaps the Newtonian law
of the inverse square {}``is inexact when applied on short distances''
\cite{layzer}. It is with the release of Einstein's theory of general
relativity (TGR) in 1915 when our conception of gravitational phenomena
radically changed. This theory is able to account for the discrepancies
that we have been talking about, particularly, the one which turned
out to be one of its most notorious predictions: Mercury's perihelionshift.
Additionally, TGR demonstrated that, in the low velocity limit relative
to the speed of light and in presence of weak gravitational fields,
it was equivalent to Newton's gravitation theory. 

It is important to remark that the new theory, for the sake of a covariant
expression (universal principle of covariance) of physical laws, does
not use Newton's formulation based on forces, which arbitrarily depends
on the selected coordinate system to describe the phenomenon, but
instead uses tensor field equations which expressions are independent
from the chosen coordinate systems. In this way, this equations, called
Einstein's Field Equations (EFE), satisfy the theoretical need to
express a covariant law of gravitation in the TGR. As we know, in
tensor notation and conventional units, we can write them as, \begin{equation}
G_{\mu\nu}=\frac{8\pi G}{c^{2}}T_{\mu\nu}\label{21}\end{equation}
 where $G_{\mu\nu}$ is Einstein's tensor, $c$ the light's speed
and $T_{\mu\nu}$ is the energy-momentum tensor. The way this equation
relates geometry (left side) with physics (right side) is the reason
why the TGR is often called {}``geometrodynamics''. However, to
get the EFE's solution means to solve, in the general case, ten second-order
partial differential equations, a tiresome work even with today's
constant advances in the computation processors' technology. This
is an important reason because, unless physical conditions demands
TGR (speeds close to the speed of light and strong gravitational fields),
Newton's gravitation law is still commonly used as a perfectly valid
theory. 

In recent years the exploration of non-relativistics schemes to calculate
astrodynamics effects has become more and more active (see for example
the references of Wang\'{}s quantum-corrected newtonian force \cite{Wang},
the Calura et al. post-Newtonian planetary equations \cite{Calura}
or the Jefimenko\'{}s Gravito-Cogravitism theory \cite{Jefi}). However,
it is important to remind that it is indeed possible \cite{weinberg},\cite{dinverno},\cite{ABS}
to generalize the notion of force, originally defined in a 3d-space,
in order to be able to use it in TGR's 4d-space (Minkowski' s force).
For the case of the obtention of a gravitational force generalized
to a 4d-space that includes relativistic effects, Adler, Bazin \&
Schiffer \cite{ABS2} and Weinberg \cite{weinberg2} propose interesting
procedures worth consulting. \ However, it is also possible to find
an expression for a gravitational force of this kind beginning with
a given solution of the EFE and the use of an equation which possesses
-as equation (\ref{21}) does- a geometro-dynamical character: Binet's
equation. 

In this paper we want to present a general scheme developed for obtain
classical gravitational forces from general relativistic settings.
This expressions can describe purely relativistic effects (as perihelionshift,
for example) with just the use of Newtonian theory. The application
of the procedure is briefly described in the section 2 and 3 for the
Schwarzschild case. The general method is summarized and the discussion
is presented in section 4. Finally, we are using here the MKS system.

\section{Movement of a Test Particle in a Gravitational Field}

We are interested in the dynamic study of a test particle placed in
a gravitational field. In any advanced Mechanic's text \cite{Marion},
\cite{Fowles}, \cite{Cabannes}, it is possible to find the procedure
to obtain Binet's equation: \begin{equation}
\frac{d^{2}u}{d\theta^{2}}+u=-\frac{F(u^{-1})}{\mu h^{2}u^{2}}\label{3}\end{equation}
 where $u=1/r$, $\mu=(Mm)/(M+m)$ is the reduced mass, and $h=r^{2}\dot{\theta}$
(the dot means a classical time derivative) is the angular momentum
per mass unit and $F$ is the force involved. Thus, this equation
relates the orbit's geometry (left side) to the particle's dynamics
(right side). As long as the orbit is given, the force that determines
it can be obtained and vice versa. 

Now, if $F$ is given by Newton's Universal Gravitation Law, which
means, if we study the \textbf{classical case}, we can find with (\ref{3})
that \begin{equation}
u(\theta)=\frac{A}{h^{2}}\left[1+e\cos(\theta)\right]\label{4}\end{equation}
 where $A\equiv G(M+m)$ , $e\equiv\frac{Ch^{2}}{A}$ (eccentricity)
and $C$ is an integration constant. 

This is the essence of the method we want to introduce. For the \textbf{relativistic
case}, it is first necessary to solve the EFE for certain physical
conditions. In this paper, to make it simpler, a static and spherically
symmetrical mass distribution is going to be taken as the source of
the gravitational field. Under these conditions, from (\ref{21}),
Schwarzschild' s solution \cite{sch} is obtained, and we can write
it as the following line element, \begin{equation}
ds^{2}=\left(1-\frac{2m_{g}}{r}\right)dt^{2}-\left(1-\frac{2m_{g}}{r}\right)^{-1}dr^{2}-r^{2}d\Omega^{2}\label{elemlin}\end{equation}
 where, $d\Omega^{2}\equiv d\theta^{2}+sen^{2}\theta d\phi^{2}$,
and \begin{equation}
m_{g}=\frac{GM}{c^{2}}\label{radio grav}\end{equation}
 is called geometrical mass or gravitational radius. 

From the study of the geodesics obtained from (\ref{elemlin}) it
results that the trajectory followed by the test particle in the discussed
gravity field is given, using a first-order perturbation method and
after some simplifications, by \cite{dinverno}\begin{equation}
\widetilde{u}(\theta)\cong\frac{m_{g}c^{2}}{\tilde{h}^{2}}\left\{ 1+e\cos\left[\theta\left(1-\epsilon\right)\right]\right\} \label{7}\end{equation}
 in non-geometrized units ,where, \begin{equation}
\epsilon=3\frac{m_{g}^{2}c^{2}}{\tilde{h}^{2}}\label{8}\end{equation}
 is the perturbation parameter, a very small quantity, and $\tilde{h}$
is the relativistic angular momentum per mass unit. 

Note the similarity between equations (\ref{7}) and (\ref{4}). From
this relativistic result Mercury's perihelionshift can be obtained,
whereas with (\ref{4}) this effect does not appear. Note also that
(\ref{7}) must be equal to (\ref{4}) when $\epsilon=0$ , obtaining
this way the relation between the classical and relativistic momentum:
\begin{equation}
\tilde{h}=\sqrt{\frac{M}{M+m}}h\label{9}\end{equation}

\section{Gravitational Force from the Schwarzschild\'{}s Metric}

Having already noticed that equation (\ref{7}) contains relativistic
information (in fact, it is a solution of the EFE) and that it is
precisely an orbital equation, we can, from a mathematical point of
view, introduce it into Binet' s equation (\ref{3}), obtain the corresponding
force and study the physical consequences of it all. 

We would like to know how should the force's expression be on (\ref{3})
in order for (\ref{7}) to become its solution. Thus, let us rewrite
(\ref{3}) in this way: \begin{equation}
\frac{d^{2}\widetilde{u}}{d\theta^{2}}+\widetilde{u}=-\frac{\tilde{F}(\widetilde{u}^{-1})}{\tilde{\mu}h^{2}\widetilde{u}^{2}}\label{10}\end{equation}
 where $\tilde{u}=1/r$ and $\tilde{\mu}$ has been written instead
of $\mu$ (the reduced mass) because we do not know if, in such a
procedure, the definition of reduced mass remains unaltered. 

Noting that equation (\ref{7}) involve a first-order perturbation
method, from (\ref{8}), (\ref{9}) and (\ref{10}), after tedious
calculations and neglecting second order terms in $\epsilon$, we
obtain: \begin{eqnarray}
\tilde{F} & = & -\frac{G(M+m)(6m_{g}+r)}{r^{3}}\tilde{\mu}\nonumber \\
 &  & +\left[\frac{G(M+m)(3m_{g}+2r)}{r^{3}}\tilde{\mu}\right]\epsilon\label{fbar}\end{eqnarray}

We must now require that as long as $\epsilon\rightarrow0$ , the
force $\tilde{F}\rightarrow-G\frac{Mm}{r^{2}}$ ; doing this we can
find that \begin{equation}
\tilde{\mu}=\frac{Mmr}{(M+m)(6m_{g}+r)}\label{mu}\end{equation}
 which, as we shall see, is some kind of \ {}``generalized reduced
mass{}``. 

If we now substitute (\ref{mu}) in (\ref{fbar}), the general first-order
gravitational force for the Schwarzschild's solution is obtained: 

\begin{equation}
\tilde{F}=-G\frac{Mm}{r^{2}}+3\frac{m_{g}^{2}c^{2}GMm\left(2c^{2}r+3MG\right)}{\tilde{h}^{2}r^{2}\left(c^{2}r+6MG\right)}\label{fufull}\end{equation}

We can see in the second term of (\ref{fufull}) the relativistic
angular momentum per mass unit $\tilde{h}=r^{2}\dot{\theta}$ (here
the point indicates a derivative with respect of proper-time). Remember
that for conservative systems where the angular momentum is conserved,
$\tilde{h}$ is constant. If this is the case, we can still rewrite
(\ref{fufull}); for example, note that if very low speeds related
to the speed of light, $c,$ are considered (as in the Solar System),
it is possible to write: \begin{equation}
\tilde{h}\cong Vr\end{equation}
 where $V$ is the orbital velocity of the test particle. We then
obtain, \begin{equation}
\tilde{F}=-G\frac{Mm}{r^{2}}+3\frac{G^{3}M^{3}m\left(2c^{2}r+3MG\right)}{c^{2}r^{4}V^{2}\left(c^{2}r+6MG\right)}\label{15}\end{equation}

If we wish to express the force $\tilde{F}$ in (\ref{15}) as a function
of $r$ alone, without the orbital velocity $V$ of the test particle,
we proceed as in the classical scheme: given that we are working with
a classical force, we assume Newton's laws are valid for it, and we
make (\ref{15}) equal to the centrifuge force \begin{equation}
F_{c}=m\frac{V^{2}}{r}\end{equation}
 obtaining, \begin{equation}
V_{1}=\pm\frac{c}{6rD}\sqrt{3D\left(D-r\right)\left(Dr+\sqrt{Dr\left(3r^{2}-Dr-D^{2}\right)}\right)}\label{vel1}\end{equation}
 and \begin{equation}
V_{2}=\pm\frac{c}{6rD}\sqrt{3D\left(D-r\right)\left(Dr-\sqrt{Dr\left(3r^{2}-Dr-D^{2}\right)}\right)}\label{vel2}\end{equation}
 where $D=r+6m_{g}$. 

Substituting any of these expressions in (\ref{15}) we can write
$\tilde{F}$ as a function of $r$ alone.

\section{Analysis and Conclusions}

We have presented the procedure directly applied to the Schwarszchild
solution. It should be convenient to summarize the general method
as follows: 

\begin{enumerate}
\item Select an arbitrary spacetime. The line element must be completely
determined. 
\item Obtain the radial equation of motion for a test particle via the corresponding
geodesic equation. Note the order of the perturbation method(s) involved,
if any. 
\item Introduce the equation of motion in the Binet\'{}s equation, (\ref{10}).
The corresponding force should be writen in terms of the perturbation
parameter(s). Check out the concordance with the order of the perturbation(s). 
\item Require that as the perturbation parameter(s) tends to zero, the preliminar
force tends to Newtonian force. This condition provides the $\tilde{\mu}$
factor$.$
\item Write the resulting expression only in terms of classical variables. 
\end{enumerate}
With the application of this method, we have derived the expression
(\ref{15}), the Schwarzschild\'{}s force from now on, including
relativistic effects and which, in strong fields, differs from Newton's
Law of Universal Gravitation. Figure 1 shows this difference for a
Sun-Mercury-like system at very small distances from the source. Note
that this is in accord with Newcomb\'{}s idea \cite{layzer}. Certainly,
when we come near the source, the field's intensity increases, and
it is precisely at this moment when the relativistic effects contained
in the second term of the Schwarzschild\'{}s force are notorious.
The asymptotic behaviour of the expression (\ref{15}) is in perfect
agreement with the predictions of Newton's gravitational law (\ref{Fu})
as it might have been expected. 

The Schwarzschild\'{}s force, is also in concordance with the predicted
general relativity correction to Newtonian gravitational motion (see
references \cite{ABS2}, \cite{Golds}; the $r^{-3}$ corrective potential
can be derived from eq. (25.42) in reference \cite{Mis}). In fact,
for weak fields, but still non Newtonian fields (for example, in the
Sun case $m_{g}=1.4766$ $km$), we have \begin{equation}
\frac{G^{3}M^{3}m\left(2c^{2}r+3MG\right)}{c^{2}r^{4}V^{2}\left(c^{2}r+6MG\right)}\simeq2\frac{G^{3}M^{3}m}{c^{2}r^{4}V^{2}}\end{equation}
 and the Schwarzschild Force has an explicit $r^{-4}$ term. 

Naturally, from (\ref{15}) is possible to find the perihelionshift
advance, so it can help to show, just via classical mechanics, this
important result of the TGR. Bearing this in mind, it is also illustrative
to compare the Schwarzschild\'{}s force with the results of the Gravito-Cogravitism
approach to find the periastron advance \cite{Matos}: the total gravito-cogravitational
force has a roughly similar $r^{-4}$ dependent term. 

Figure 2 shows the comparisson between the classical velocity and
the obtained expressions (\ref{vel1}) and (\ref{vel2}). It is important
to remark that for short distances velocities $V_{1}$ and $V_{2}$
are slower than the Newtonian one, but asimptotically agrees with
the former. This is due to the fact that the second term in equation
(\ref{15}) is important when the field is strong, making the Schwarzschild\'{}s
force $\tilde{F}$ less intense than the classical force (\ref{Fu}),
as seen in Figure 1; therefore, the equilibrium condition between
gravitational and centrifugal force demands equations (\ref{vel1})
and (\ref{vel2}) to be lesser than the classical orbital velocity.
As the field becomes weaker, the Schwarzschild\'{}s force matchs
the Newton\'{}s force, and there is not \textit{a priori} way to
distinguish between the velocities. Note that the velocities matching
occurs for diferents distances. Indeed, for a Sun-Mercury-like system,
as the one presented in Figure 2, $V_{1}$ coincides with Newtonian
orbital velocity for $r_{1}\gtrsim$ $4x10^{2}Km$ from the source,
while $V_{2}$ coincides for $r_{2}\gtrsim5x10^{5}Km$. 

It is an interesting issue that there are \textit{two} velocities,
(\ref{vel1}) and (\ref{vel2}), that satiesfies the equilibrium condition
between centrifugal and gravitational forces instead of one. We report
that as a curiosity of the ansatz we have used since we had not find
any physical reason to discard equation (\ref{vel2}). As far as we
know this is a new result, and, since this bifurcation occurs for
very near distances from the source it could be quite difficult to
prove the existence (or not) of this effect. 

On the other hand, we also have obtained the interesting equation
(\ref{mu}) that is a mass depending on the separation between the
interacting bodies and which, in the weak fields limit (small $m_{g}$),
turns into the known expression for the reduced mass of a two bodies
system. Also observe that the same asymptotic behaviour occurs when
the bodies' separation is quite large (Figure 3). The origin of (\ref{mu})
can be attributed to the fact that the effects spawned from space-time
curvature in TGR have been transfered, into the present scheme, in
the variation of this generalized mass with respect to the interacting
bodies' distances. 

As we have mentioned at the introduction, equation (\ref{10}) possesses
a geometro-dynamical character in the sense that it relates the test
particle's trajectory (geometry) to the force (dynamics) acting on
it. Note too that $\tilde{\mu}$ is present in the denominator of
the dynamical member of (\ref{10}): from this we can deduce that
as long as we approach the source, the effects on the geometrical
member become increasingly notorious, meaning that the trajectory
will progressively differ from the one predicted by Newton's law of
gravitation. Examples of this are Mercury and Icarus because of their
nearness to the Sun. 

Finally, we want to point out that more detailed future investigations
are necessary to explore all the wealth of this ansatz; this method
possesses a great deal of conceptual value, because it shows the way
to describe purely relativistic effects through the use of a completely
classical scheme. Moreover, it is general: given a trajectory predicted
in a relativistic setting, it can be obtained a corresponding gravitational
force {}``only'' expressed in classical notions. 

\begin{center}\textbf{Acknowledgments}\end{center}

The authors want to thank the invaluable help of Professor Zigor Mu\~{n}oz
and Mr. Javier De La Hera in the preparation of the paper. Also, the
stimulating comments of Dr. Jos\'{e} Ferm\'{\i}n showed helpful
in the development of the investigation.

\newpage{}\textbf{Captions.}

Figure 1. Newtonian (solid) and obtained (dots) forces. 

Figure 2. Classical velocity (dotdashed) and expressions (\ref{vel1}),
with solid line, and (\ref{vel2}) with dots. 

Figure 3. Reduced mass (solid line) and $\tilde{\mu}$ parameter (dots
line). 
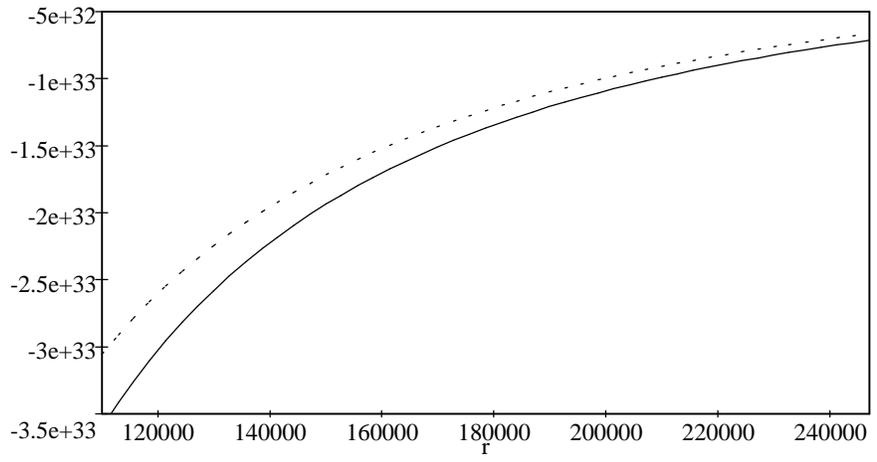
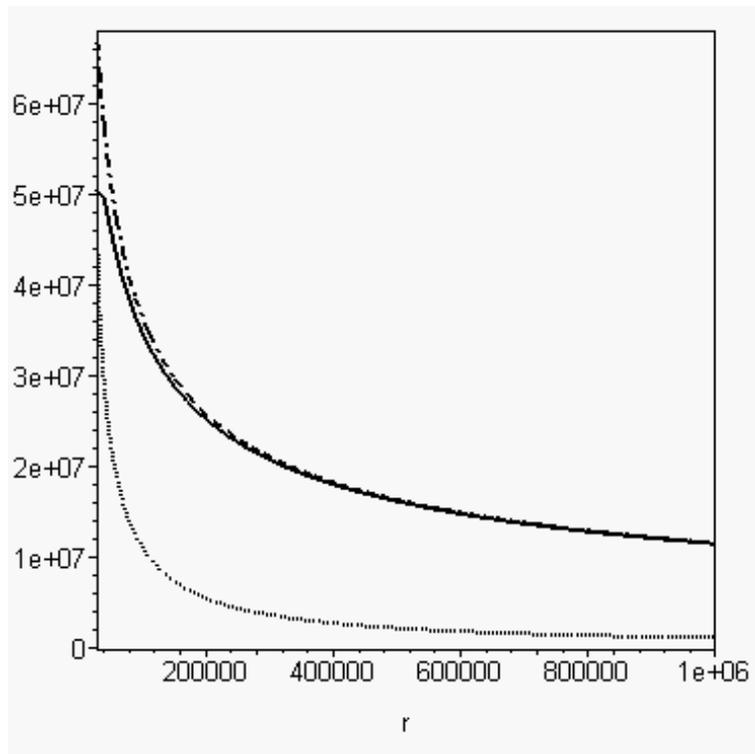
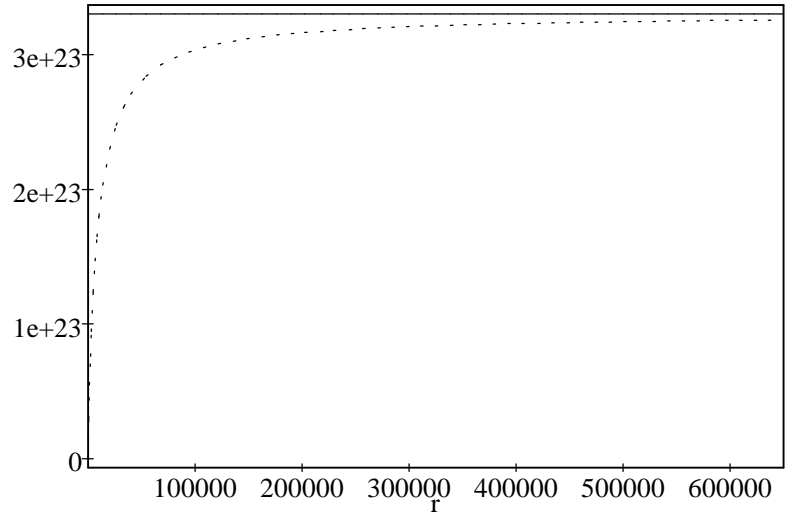
\end{document}